\title{Anisotropic representations for E(3)-equivariant machine learning coarse-grained potentials}
\author{Varun Shankar and Emil Annevelink}
\date{}

\documentclass[11pt, letterpaper]{article}
\usepackage[utf8]{inputenc}
\usepackage[T1]{fontenc}
\usepackage{graphicx} 
\usepackage{fancyhdr} 
\usepackage[
    margin=1in,
    headheight=1in,
    top=1.5in,
    bottom=1in,
]{geometry} 
\usepackage{amsmath}
\usepackage[backend=biber,style=numeric,sorting=none,doi=true,url=true,eprint=false]{biblatex}
\AtEveryBibitem{%
  \iffieldundef{doi}{}{\clearfield{url}}%
}

\newcommand{\logoimage}{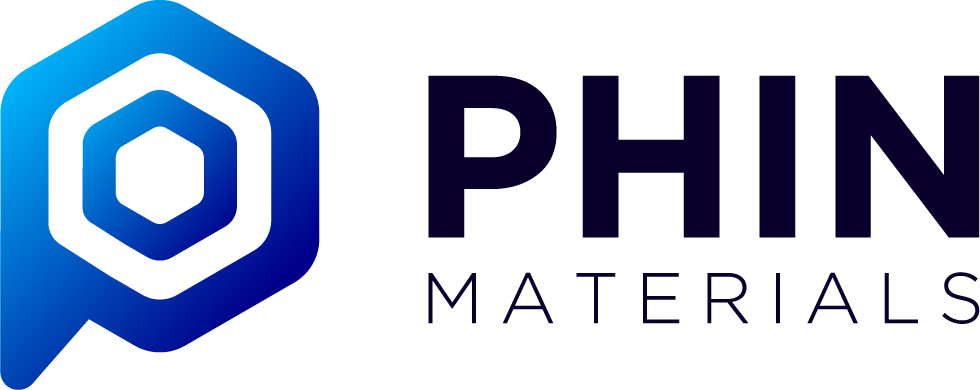} 
\newcommand{\logoheight}{1.5cm} 

\fancypagestyle{plain}{
    \fancyhf{} 
}

\begin{document}

\maketitle
\thispagestyle{fancy}
\vspace{3cm} 
\pagenumbering{arabic}
\section*{Executive Summary}
Coarse-graining (CG) lowers the computational cost of atomistic simulations by representing groups of atoms as effective interaction sites, reducing the degrees of freedom of the system but often compromising structural fidelity or requiring system-specific parameterization. Here, we introduce a novel anisotropic machine learning CG potential that extends the point particle representation of atomic nuclei to massive ellipsoidal beads with orientation-dependent features, enabling the learning of energies, forces, and torques directly from atomistic data. The anisotropic representation is physically motivated for polar and asymmetric molecules, where directional interactions and shape anisotropy play important roles in determining structure and dynamics. Using an equivariant message-passing neural network, the model accurately reproduces radial and angular distribution functions as well as relative orientation correlations in liquid water, demonstrating that both translational and rotational dynamics are well captured. Comparison with an isotropic baseline reveals that the lack of orientation information leads to systematic errors in short and long range order and degradation of angular correlations, proving orientation features are essential for accurate coarse-graining. The anisotropic model also exposes rotational structural observables fundamentally inaccessible to isotropic representations, with minimal computational overhead. Even for coarse-graining just three degrees of freedom, CG simulations achieve 7-27$\times$ speedups while preserving structural fidelity, highlighting the efficiency gains of this systemic reduction. This framework establishes the feasibility and necessity of learned equivariant representations for anisotropic CG modeling and provides a path towards accurate and efficient mesoscopic simulations of complex molecular liquids, polymers, and biomolecular systems.

\pagebreak

\section{Introduction}

Atomistic simulations at density functional theory (DFT) accuracy form the basis of predictive modeling in chemistry and materials science. However, the computational cost of DFT scales steeply with system size \cite{kresse_1996}, limiting practical simulations to hundreds of atoms and short timescales. Many processes of interest, such as polymerization, phase separation, or biomolecular kinetics, occur at mesoscopic scales involving millions of atoms and microsecond dynamics. Classical analytical force fields can reach these scales but sacrifice the accuracy required for predictive modeling of complex chemical phenomena. In order to digitize materials development and accelerate discovery, modeling approaches that extend the cost-accuracy Pareto front to achieve quantum-level fidelity at mesoscopic scales must continue to be developed.

Over the last several years, machine learning interatomic potentials (MLIPs) have emerged to bridge this gap \cite{deringer_2019}. Methods such as Gaussian approximation potentials, descriptor neural networks, and equivariant graph networks have demonstrated that near-DFT accuracy can be achieved at orders-of-magnitude lower cost \cite{kocer_2022}, and MLIPs are now widely applied across catalysis, solid-state materials, and molecular liquids \cite{wines_2025,Batatia2022mace,wood_2025,rhodes_2025}. However, MLIPs remain limited to tens of thousands of atoms due to their increased cost over analytical force fields, preventing access to the scales required for mesoscopic phenomena. The fundamental barrier is the number of degrees of freedom, as MLIPs retain full atomistic resolution.

Coarse-graining (CG) methods address this limitation by reducing the DoFs, replacing atomistic interactions with effective mean potentials that represent groups of atoms as single interaction sites \cite{joshi_2021}. This enables mesoscale simulations at computational cost comparable to classical force fields while aiming to preserve the structural fidelity of underlying reference data. Traditional CG models, however, rely on rigid parameterization strategies that often sacrifice accuracy or transferability. Top-down approaches are optimized to reproduce thermodynamic observables such as density or compressibility, while bottom-up approaches derive effective potentials from atomistic data. In both cases, trade-offs arise: reduced computational burden can come at the expense of structural fidelity, and fitting models can be a time-consuming, expensive process that requires iterative optimization and careful attention from researchers. These challenges have motivated ongoing efforts to develop more flexible and general CG models.

The success of MLIPs at the atomistic scale naturally raises the question of whether similar data-driven strategies can be extended to the coarse-grained domain. If MLIPs have transformed atomistic modeling by combining accuracy with efficiency, then machine learning coarse-grained potentials (MLCGPs) represent the next logical step to reach mesoscopic scales. MLCGPs aim to replace traditional parameterized CG models with flexible, data-driven potentials capable of capturing both thermodynamic and structural fidelity, while retaining the computational efficiency of coarse-grained simulation. A particularly promising direction is the use of anisotropic particles as coarse-grained sites, which provide richer representations than isotropic beads by accounting for features such as polarity, chirality, and shape anisotropy. For many molecular systems, an anisotropic representation is more physically motivated and can reduce information loss during coarse-graining.

In this work, we combine E(3)-equivariant graph neural networks with explicit orientation degrees of freedom to develop a MLCGP that learns anisotropic representations of local environments containing ellipsoidal particles with shape and orientation. Using liquid water as a test case, we assess the quality of the MLCGP by evaluating its ability to reproduce structural observables from atomistic reference data and directly compare this anisotropic model against an isotropic baseline to determine the effect of orientation information on structural accuracy.

\section{Background}

The core principle of coarse‑graining is to reduce the number of degrees of freedom in a system by mapping groups of atoms onto effective interaction sites or beads. This mapping smooths the underlying potential energy surface, suppressing high‑frequency fluctuations and enabling simulations that reach mesoscopic length and time scales. While the loss of atomistic detail is unavoidable, the hope is that essential physics such as structure, thermodynamics, and relevant collective dynamics can be preserved in the reduced representation. The effectiveness of any CG scheme therefore depends not only on the quality of the mapping, but also on the form of the effective interactions that govern the coarse‑grained sites \cite{Noid2013_CGmodels}.

A broad class of models has been developed for this purpose. Isotropic bead‑based models, such as the Martini force field, have been widely adopted for biomolecular and soft‑matter systems, providing tractable models for proteins, lipids, and polymers. Martini models map several heavy atoms to various bead‑types, governed by Lennard‑Jones (LJ) potentials that are calibrated through top‑down and bottom‑up approaches \cite{Marrink2023_MartiniReview}. Dissipative particle dynamics (DPD) has proven useful for mesoscopic fluids by embedding hydrodynamic behavior into a stochastic coarse‑grained framework. DPD is a particle‑based method that adds dissipative and stochastic forces to account for unresolved DoFs to the typical conservative forces used in atomistic simulations \cite{Espanol2017_DPD}. While successful, these methods are usually tailored to specific systems through carefully chosen mappings and parameterizations, limiting their generalizability. Furthermore, isotropic bead approximations often fail to capture directional features of interactions such as polarity or chirality. To overcome this, anisotropic coarse‑grained sites like ellipsoids or other finite‑sized particles have long been considered \cite{GayBerne1981_ModOverlap,GayBerne1996}. These provide orientation‑dependent interactions and a richer representational capacity. Despite the more flexible modeling framework introduced by aspherical potentials, the requirement of system‑specific parameterization in CG models remains a key bottleneck.

In parallel, machine‑learning interatomic potentials (MLIPs) have transformed atomistic modeling by directly learning potential energy surfaces from quantum‑mechanical or force‑field reference data. Landmark MLIP formulations such as the Behler–Parrinello neural network potentials \cite{BehlerParrinello2007} and Gaussian Approximation Potentials (GAP) \cite{Bartok2010_GAP} demonstrated early that learned potentials can approach ab‐initio accuracy, and more recent equivariant graph neural networks have improved data efficiency and generalization \cite{Batzner2022_NequiP}. Importantly, MLIPs have also introduced methodological advances that are highly relevant to coarse‑graining. Active learning frameworks allow the model to adaptively select training data from regions of configuration space where predictions are likely to be inaccurate, dramatically improving development efficiency \cite{Smith2022_ActiveLearning}. Related to active learning, uncertainty quantification (UQ) techniques, such as ensembles and Bayesian methods, have provided practical estimates of model reliability, which can be integrated into automated training and active‑learning pipelines. Such methods have already demonstrated that uncertainty‑biased dynamics can accelerate discovery of informative configurations for training MLIPs \cite{Li2023_UQ_MLIP}. These developments have helped establish MLIPs not only as accurate models but also as part of a scalable workflow for automated potential construction.

The success of MLIPs has motivated a flurry of recent activity in developing CG potentials with machine learning \cite{Sultan2023_CGReview}. Several works have adapted existing MLIP architectures to the CG domain simply by targeting force‑matching mean forces on beads \cite{Zhang2020_DeePCG,Husic2023_CGSchNet}. Active learning has been successfully applied to address the data efficiency challenge in MLCGPs. Loeffler et al. demonstrated that active learning using nested ensemble Monte Carlo could train a descriptor-based CG potential for bulk water with approximately 300 reference configurations. Notably, data quality over quanity was deemed to be critical for accuracy, indicating the importance of sampling method for dataset generation \cite{Loeffler2020_CGActive}. Duschatko et al. applied the FLARE framework \cite{vandermause_2020} to CG potentials, showing improved performance over non-active learning approaches. However, Gaussian process models scale unfavorably with dataset size and may not be suitable for all applications \cite{Duschatko2022_CGFreeEnergy}. Furthermore, equivariant networks have also been leveraged in CG‐MLPs, validating this physical prior's importance in addressing data efficiency. Loose et al. compared DeePMD (non-equivariant) \cite{zhang_2018} and Allegro (equivariant) \cite{musaelian_2023} models for CG potentials, highlighting the improved accuracy of the latter within a small-data regime \cite{Loose2023_CGEquiv}. These approaches leverage spherical potentials and predict the energies and forces of beads. However, orientation and torque, which are required for anisotropic CG models, remain underrepresented in current ML frameworks. Recent work on anisotropic coarse‑graining has shown that mean force- and torque‑matching formulations can reproduce both forces and angular dynamics in molecular liquids. Wilson et al. employed a descriptor-based network that encodes local environments using symmetry functions, accounting for both position and orientation, to learn an anisotropic CG potential. While the model demonstrated good accuracy on structural and thermodynamic observables for two distinct molecular systems, ML potentials have largely gravitated towards learned representations rather than predefined descriptors for improved accuracy and generalization \cite{Wilson2023_AnisoCG,kocer_2022}. 

Despite this progress, two key limitations remain. Anisotropic CG approaches have not been combined with equivariant message-passing networks that use learned representations, as prior anisotropic work relies on hand‑crafted descriptors and equivariant networks have been applied only to isotropic beads. Furthermore, the added complexity of orientation features has not been systematically evaluated through direct comparison with isotropic baselines, leaving their practical benefit unclear for polar and anisotropic molecular systems.

\section{Research Focus}

This work addresses the gaps in existing MLCGP literature by combining E(3)‑equivariant graph neural network architectures with explicit orientation inputs, extending learned representations to anisotropic particles that predict torques in addition to forces. To quantify the benefit of orientation information, we perform a direct comparison against an isotropic variant trained on identical data, isolating the effect of anisotropic features on structural accuracy.

Each molecule is mapped to an ellipsoidal bead through a rigid body transformation based on the inertia tensor, defining shape and orientation at each timestep. The model predicts bead‑level energies, forces, and torques from atomistic reference trajectories, with node features constructed from both shape parameters and spherical harmonic projections of principal axes. Structural observables from both the anisotropic and isotropic potential formulations are compared against coarse‑grained all‑atom reference data for liquid water.

Results show that the anisotropic model quantitatively matches the reference data across radial, angular, and orientation distributions, while the isotropic variant exhibits systematic errors in nearest‑neighbor structure and pronounced deviations in angular correlations at intermediate length scales. These differences demonstrate that orientation‑dependent features are not simply refinements but necessary components for capturing local geometry in systems governed by directional interactions. The findings establish both the technical feasibility of combining equivariance with orientational degrees of freedom and the tangible benefit of doing so for polar molecular systems.

\section{Methods}

\subsection{Coarse-Grained Mapping}

Each molecule in the atomistic system is represented as a single anisotropic coarse-grained particle by fitting a uniform ellipsoid to its atomic mass distribution. Given atomic coordinates $\mathbf{r}_i$ and masses $m_i$, the molecular center of mass (COM) is computed as
\begin{equation}
\mathbf{r}_{\mathrm{COM}} = \frac{\sum_i m_i \mathbf{r}_i}{\sum_i m_i}.
\label{eq:com}
\end{equation}
Atomic coordinates are shifted to the COM frame, $\mathbf{r}'_i = \mathbf{r}_i - \mathbf{r}_{\mathrm{COM}}$, and used to construct the inertia tensor,
\begin{equation}
\mathbf{I} = \sum_i m_i \left( \|\mathbf{r}'_i\|^2 \mathbf{I}_3 - \mathbf{r}'_i \otimes \mathbf{r}'_i \right),
\label{eq:inertia}
\end{equation}
whose eigenvalues $(I_a, I_b, I_c)$ and eigenvectors $(\mathbf{e}_1, \mathbf{e}_2, \mathbf{e}_3)$ define the principal moments and axes of rotation, respectively. The matrix $\mathbf{R} = [\mathbf{e}_1, \mathbf{e}_2, \mathbf{e}_3]$, representing a rotation to the local coordinate frame, is enforced to satisfy $\det(\mathbf{R}) = 1$, ensuring consistent right-handed coordinate frames across beads. The corresponding ellipsoid semi-axis magnitudes $(a, b, c)$ are determined by solving the following system of equations given by the standard inertia relations for a uniform-density ellipsoid of total mass $M$:
\begin{equation}
I_a = \frac{M}{5}(b^2 + c^2), \quad
I_b = \frac{M}{5}(a^2 + c^2), \quad
I_c = \frac{M}{5}(a^2 + b^2).
\label{eq:axes}
\end{equation}
The resulting mapping defines for each bead a center-of-mass position $\mathbf{r}_{\mathrm{COM}}$, semi-axes $(a, b, c)$, and orientation represented by a quaternion $\mathbf{q}$.

\begin{figure}
    \centering
    \includegraphics[width=0.9\linewidth]{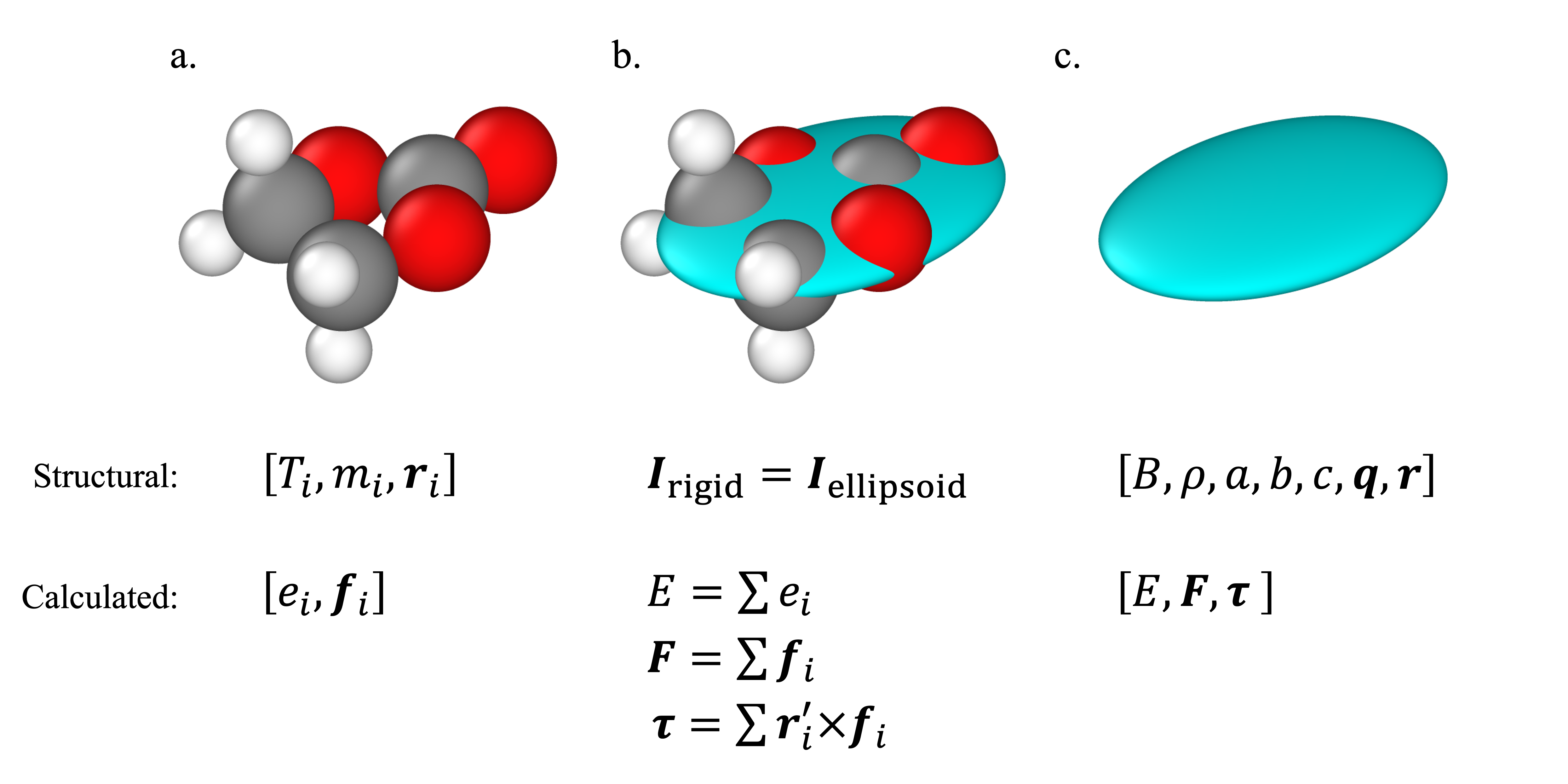}
    \caption{Presents an overview of the coarse-graining procedure for an ethylene carbonate molecule. \textbf{a.} The atomic representation defines for each atom a type $T_i$, mass $m_i$, and position $\mathbf{r}_i$, with computed values for energy $e_i$ and forces $\mathbf{f}_i$. \textbf{b.} An ellipsoid is fit to each molecule in a rigid body mapping by ensuring inertial consistency, while computed fields are aggregated appropriately. \textbf{c.} The beaded representation defines a bead type $B$, density $\rho$, semi-axes $(a,b,c)$, orientation $\mathbf{q}$, and COM position $\mathbf{r}$, with bead-level properties energy $E$, forces $\mathbf{F}$, and torques $\boldsymbol{\tau}$.}
    \label{fig:mapping}
\end{figure}

Per-bead energies, forces, and torques are computed by aggregating atomic quantities within each mapped group:
\begin{equation}
E_{\mathrm{bead}} = \sum_i e_i, \qquad
\mathbf{F}_{\mathrm{bead}} = \sum_i \mathbf{f}_i, \qquad
\boldsymbol{\tau}_{\mathrm{bead}} = \sum_i (\mathbf{r}'_i \times \mathbf{f}_i),
\label{eq:force_torque}
\end{equation}
while the global total energy and stress remains unchanged. These quantities were used directly as supervised training targets. 
Figure \ref{fig:mapping} illustrates the ellipsoid fitting and orientation mapping process.

\subsection{Node Embedding}

Traditional MLIPs typically represent each atom as a scalar-valued embedding based solely on its chemical species and, in some cases, geometric descriptors of its local environment. In the present work, this concept is extended to anisotropic CG particles by incorporating both shape and orientation into the node representation.

For the anisotropic model (MLCGP), the node embedding for bead $i$ consists of its ellipsoidal semi-axes $(a_i, b_i, c_i)$ and an orientation-dependent component derived from its quaternion $\mathbf{q}_i$. The quaternion is first converted to the corresponding rotation matrix $\mathbf{R}(\mathbf{q}_i)$, from which the unit-normalized principal axes $(\mathbf{a}_i, \mathbf{b}_i, \mathbf{c}_i)$ in the global frame are extracted. Only two axes are required to fully define the local frame, as the third can be inferred via orthogonality and a right-handed coordinate frame. These axes are then projected onto a real spherical harmonic basis up to $l=2$, producing an equivariant set of orientation features. The complete node embedding is therefore
\begin{equation}
\mathbf{h}_i = [\,a_i, b_i, c_i, Y^{l}_m(\mathbf{a}_i), Y^{l}_m(\mathbf{b}_i))\,],
\label{eq:embedding}
\end{equation}
which captures both the shape anisotropy and the orientation of each bead. Empirically, this node-centric embedding produced more stable and generalizable representations than edge-based relative orientation encodings, such as relative rotations between pairs of beads or projections of ellipsoid semi-axes onto relative position vectors.

To demonstrate the value of anisotropy, we also trained an isotropic baseline (MLCGP-ISO) where the node embedding excludes the orientation- and shape-dependent features, resulting in a relatively vanilla isotropic equivariant ML potential centered on bead COMs. This removes all explicit orientation information from the model input while maintaining the same network architecture and training procedure, enabling comparison of model performance with and without orientation features.

\subsection{Equivariant Model Architecture}

The MLCGP employs an E(3)-equivariant message-passing neural network architecture, similar to several existing MLIP architectures \cite{Batzner2022_NequiP}. Each node carries learned features and fixed attributes, such as chemical identity, and for CG particles, anisotropic shape and orientation. Edge features contain transformations of a relative position vector $(\mathbf{r}_{ij})$, including basis-projected magnitude and spherical harmonic projections. These features are represented in their O(3) irreducible representations (scalars, vectors, and higher-order tensors) that transform equivariantly under rotation. At each interaction block, node and edge features are combined through learned tensor products that mix scalar and tensor channels while preserving overall equivariance.

The message-passing procedure iteratively updates node embeddings by aggregating information from neighboring beads based on relative displacements and orientations. The final per-node latent states are projected to scalar energy contributions $E_i$, which sum to the total potential energy,
\begin{equation}
E_{\mathrm{total}} = \sum_i E_i.
\label{eq:energy_sum}
\end{equation}
Forces and torques are obtained via automatic differentiation of this energy with respect to translational and rotational degrees of freedom, ensuring strict energy conservation.

\subsection{Force and Torque Computation}

Forces are computed directly from the energy gradient,
\begin{equation}
\mathbf{F}_i = -\frac{\partial E_{\mathrm{total}}}{\partial \mathbf{r}_i},
\label{eq:forces}
\end{equation}
while torques are obtained from the energy derivative with respect to quaternion coordinates. The quaternion gradient $-\partial E / \partial \mathbf{q}$ is converted to a Cartesian torque vector through the transformation
\begin{equation}
\boldsymbol{\tau}_i = -\mathbf{T}(\mathbf{q}_i)\frac{\partial E_{\mathrm{total}}}{\partial \mathbf{q}_i},
\label{eq:torque}
\end{equation}
where $\mathbf{T}(\mathbf{q})$ is defined as
\begin{equation}
\mathbf{T}(\mathbf{q}) =
\frac{1}{2}
\begin{bmatrix}
 -q_x & q_w & -q_z & q_y \\
 -q_y & q_z & q_w & -q_x \\
 -q_z & -q_y & q_x & q_w
\end{bmatrix}.
\label{eq:Tmat}
\end{equation}
This formulation ensures that both translational and rotational responses are energy-conserving, extending the standard MLIP autodifferentiation workflow to handle anisotropic particles.

\subsection{Active Learning Framework}

Model training was embedded within an iterative active learning (AL) loop designed to efficiently explore CG configuration space and automatically refine the dataset. The AL cycle proceeds as follows.

\begin{enumerate}
    \item \textbf{Initialization:} The process begins with a seed dataset generated from equilibrated atomistic simulations. Each configuration is mapped to its CG representation and labeled by a CG oracle, which provides bead-level energies, forces, and torques derived from high-fidelity atomistic MLIPs or reference force fields.
    \item \textbf{Training:} The MLCGP is trained on this labeled dataset using the configuration detailed below. 
    \item \textbf{Sampling:} The trained MLCGP is used to perform CG molecular dynamics simulations to sample new configurations representative of the learned potential. High error configurations are identified and marked for recalculation.
    \item \textbf{Labeling and retraining:} Selected configurations are fine-grained to an atomistic representation, evaluated by the ground truth all-atom calculator and subsequently coarse-grained, appended to the dataset, and used to retrain the MLCGP from the previous checkpoint. 
\end{enumerate}

\subsection{Training Configuration}

Training used the Adam optimizer with AMSGrad enabled, an initial learning rate of $10^{-2}$, and cosine annealing with warm restarts every 40 epochs. Early stopping was applied to the training loss with a patience of 500 epochs and a relative improvement threshold of 2\%. The loss function combined bead-level and global targets (energies, forces, torques, total energy, and stress) with coefficients scaled to normalize magnitude disparities across observables. Model performance was tracked using per-species MAE and RMSE metrics for all quantities, and the best model checkpoint was retained by minimum training loss.

\section{Results}
\subsection{Dataset and Reference Data}
\begin{figure}
    \centering
    \includegraphics[width=0.8\linewidth]{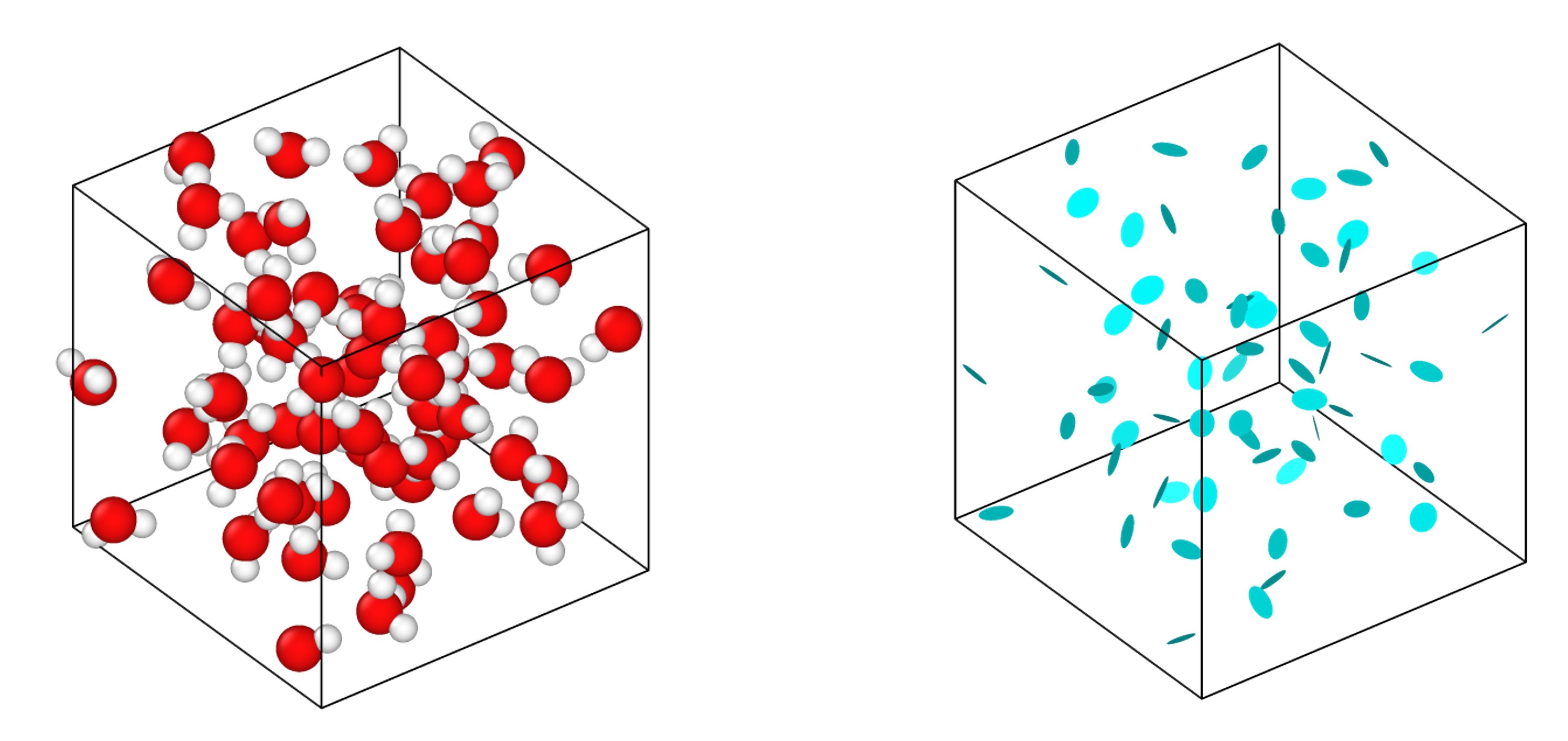}
    \caption{Initial configurations of the 64-molecule liquid water system in atomistic (left) and beaded (right) representations. Simulations were performed using the NVT ensemble at 300K for 10 ps.}
    \label{fig:initial}
\end{figure}

All results presented here were obtained for liquid water. The reference dataset consisted of a 64-molecule system simulated in the canonical ensemble (NVT) at 300 K for 10 ps. The initial structure was taken from a 100 ps NPT equilibration at 300 K and 1 atm to ensure proper density and structural relaxation. The ground truth energy and force data were generated using an all-atom (AA) molecular dynamics simulation driven by the MACE universal machine learning potential \cite{Batatia2022mace,Batatia2022Design}, which served as the ground-truth for coarse-grained training. Atomistic data were coarse-grained following the rule-based ellipsoidal mapping described earlier, providing COM positions, orientations, and corresponding mean forces and torques for each water molecule. These coarse-grained labels provided both the seed training data for the MLCGP and the reference trajectories for subsequent comparison. Figure \ref{fig:initial} shows the initial system configuration in all-atom and coarse-grained representations.

\subsection{Structural Distributions}
The fidelity of the anisotropic MLCGP was assessed by comparing structural distributions to those derived from the coarse-grained all-atom ground-truth reference (AA). To demonstrate the value of the anisotropic representation, we also compared against an isotropic baseline (MLCGP-ISO) using the same architecture and dataset, but excluding the orientation-dependent node embedding features.

\begin{figure}
    \centering
    \includegraphics[width=\linewidth]{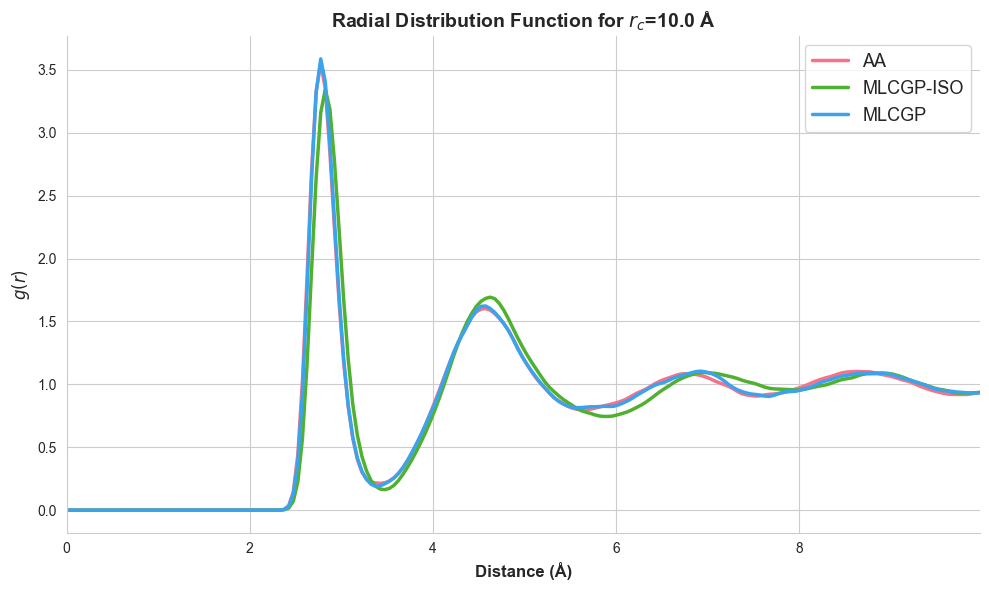}
    \caption{Radial distribution function $g(r)$ of coarse-grained bead centers for liquid water at 300K. The anisotropic MLCGP accurately reproduces the reference coarse-grained data (AA) derived from all-atom simulations using the MACE potential, with agreement in both peak positions and magnitudes. In contrast, the isotropic baseline (MLCGP-ISO) shows a clear rightward shift of the first coordination peak and reduced peak height, demonstrating that orientation features are essential for capturing correct nearest-neighbor structure.}
    \label{fig:rdf}
\end{figure}

Figure \ref{fig:rdf} shows the radial distribution function $g(r)$ of bead centers of mass. The MLCGP accurately reproduces both the position and magnitude of the first coordination peak and subsequent peaks at longer ranges, demonstrating near-perfect agreement with the reference data. In contrast, the MLCGP-ISO model exhibits a clear shift of the first and second coordination peaks, indicating that the isotropic representation fails to capture the correct nearest-neighbor distances and coordination structure. It is evident that orientation features provide additional structural information to the model that aids in reproducing postional correlations that are characteristic of the liquid water structure.

\begin{figure}
    \centering
    \includegraphics[width=\linewidth]{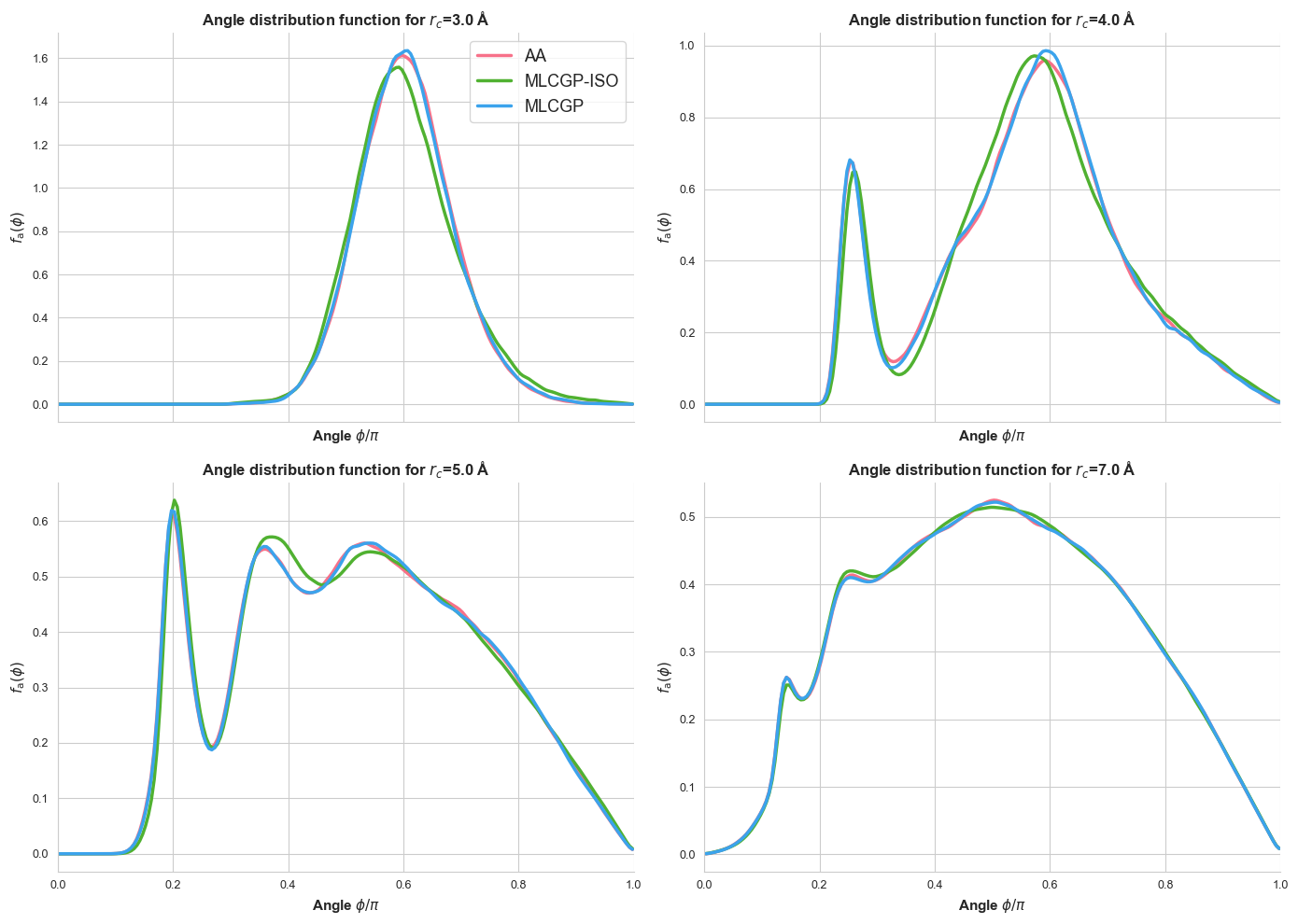}
    \caption{Angular distribution functions (ADFs) for triplets of coarse-grained beads computed at cutoff radii of 3.0, 4.0, 5.0, and 7.0\AA. The anisotropic MLCGP accurately reproduces the reference all-atom distributions (AA) across all cutoff radii. The isotropic baseline (MLCGP-ISO) shows visible discrepancies at $r_c$=3.0, 4.0, and 5.0\AA{}, where evident disagreements in peak positions and amplitudes indicate failure to capture the angular structure of the liquid.}
    \label{fig:adf}
\end{figure}

Angular distribution functions (ADFs) were computed to further probe local three-body correlations between molecular centers. As shown in Figure \ref{fig:adf}, distributions of angles between triplets of beads were evaluated at cutoff radii of 3.0, 4.0, 5.0, and 7.0\AA{}, corresponding to successively larger coordination shells. The MLCGP faithfully reproduces the characteristic features of the AA distributions across all cutoff radii, including the first-shell peak at short range and the broader multimodal structures found in longer-range correlations, demonstrating that the model correctly encodes many-body contributions to the effective potential.

Comparison with the MLCGP-ISO baseline reveals clear benefits of the anisotropic representation. At $r_c$=3.0\AA{}, the isotropic model shows visible discrepancies in the distribution. The disagreement becomes more pronounced at intermediate cutoffs ($r_c$=4.0 and 5.0\AA{}), where the MLCGP-ISO model exhibits deviations in both peak positions and amplitudes, failing to capture the complex angular structure that characterizes the first and second coordination shells. Interestingly, at $r_c$=7.0\AA{}, the isotropic model's accuracy appears to improve as angular correlations become weaker due to averaging over larger spatial regions. This pattern demonstrates that orientation-dependent features are most critical for capturing the well-defined angular structure in the first few coordination shells, where directional interactions from molecular anisotropy contribute to local geometry.

\begin{figure}
    \centering
    \includegraphics[width=0.6\linewidth]{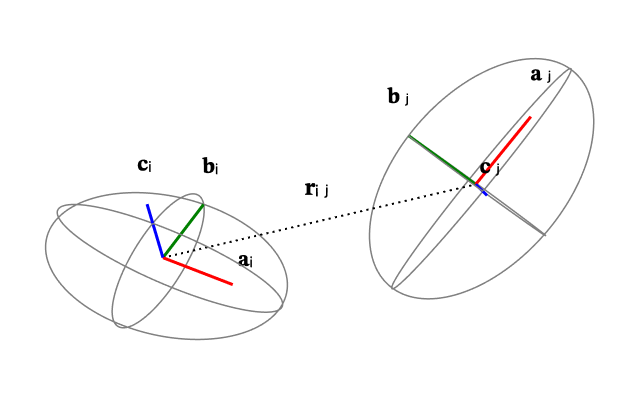}
    \caption{Depiction of a characteristic pair of ellipsoids, with local coordinate axes $(\mathbf{a}, \mathbf{b}, \mathbf{c})$ for each ellipsoid and intermolecular vector $\mathbf{r}_{ij}$ annotated. These vectors were used to compute statistical distributions of relative orientation features in order to assess anisotropic structural fidelity.}
    \label{fig:rel_orientation}
\end{figure}

\begin{figure}
    \centering
    \includegraphics[width=\linewidth]{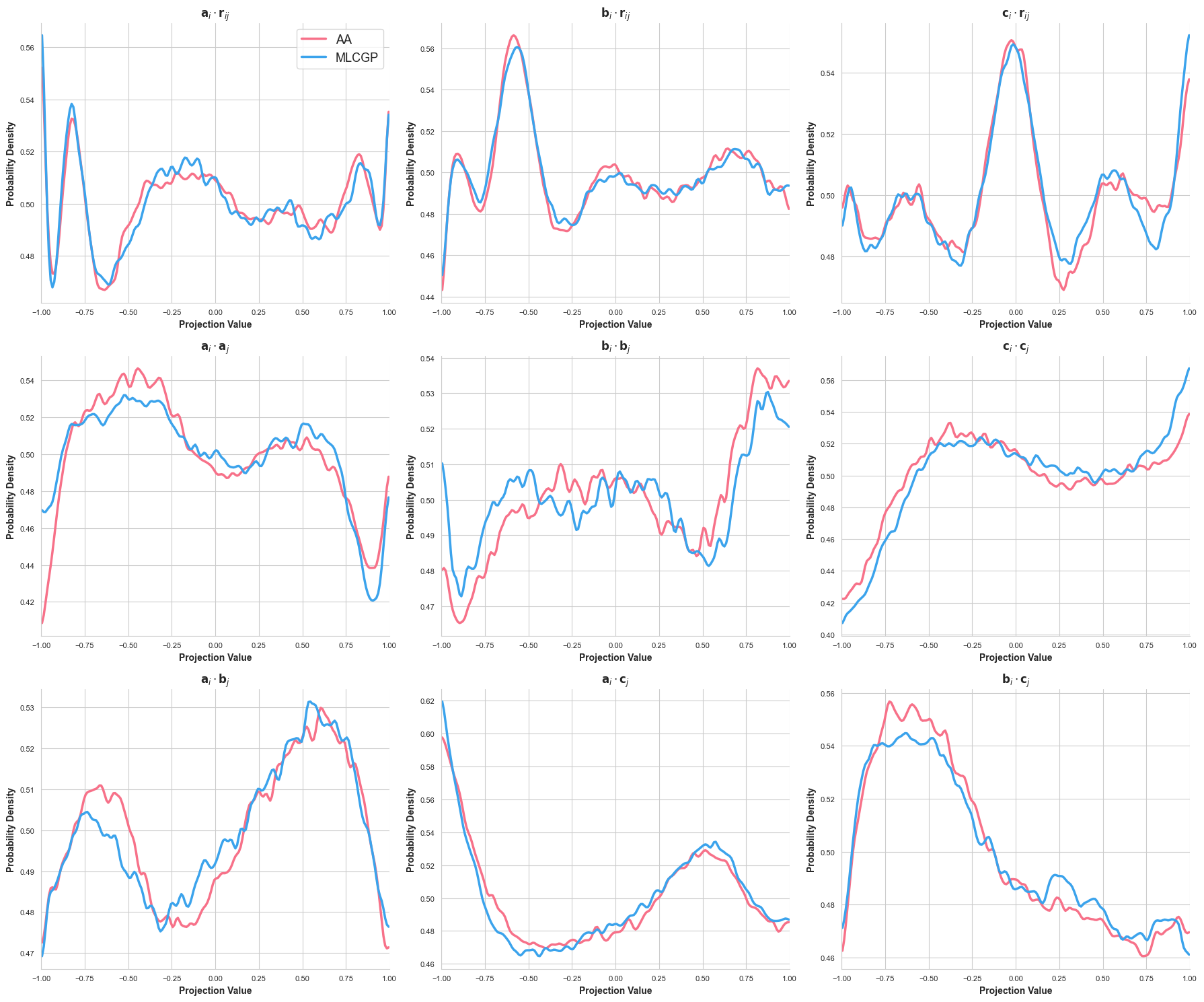}
    \caption{Distributions of relative orientation projections between neighboring coarse-grained ellipsoidal beads. These distributions include probability densities for projections of each principal axis onto the intermolecular vector and  cross-projections between corresponding molecular axes of adjacent beads. The MLCGP reproduces the orientational alignment preferences observed in the reference data, indicating preservation of the anisotropic structure of the liquid.}
    \label{fig:odf}
\end{figure}

A key advantage of the anisotropic mapping adopted here is that it enables orientation-dependent structural observables that are inaccessible in traditional isotropic bead models. Figure \ref{fig:rel_orientation} visualizes the orientation vectors used to compute statistics for a characteristic pair of beads, including projections of each principal axis onto the intermolecular vector as well as cross-projections between corresponding principal axes of adjacent beads. Figure \ref{fig:odf} presents distributions of these relative orientation projections between neighboring molecules. The resulting distributions are non-uniform, reflecting preferential orientation alignments among neighboring molecules. The anisotropic MLCGP model accurately reproduces the peak positions and magnitudes of these relative orientation correlations, demonstrating that the learned potential captures not only positional structure, but also the rotational structure of the liquid. Given that the target system consists of polar molecules, this capability proves essential for describing systems in which anisotropy influences nano- and mesoscale behavior.

Overall, the combined analysis of RDFs, ADFs, and orientation projections confirms that the anisotropic MLCGP successfully learns the effective energy landscape governing both positional and rotational correlations, achieving quantitative agreement with the reference coarse-grained dynamics derived from atomistic data. The comparison with the isotropic baseline reveals that the lack of orientation information leads to systematic errors in structural correlations. These results demonstrate that orientation-dependent features are not merely beneficial but essential for accurate coarse-graining of systems with directional interactions, validating the physical motivation for the ellipsoidal representation.

\subsection{Computational Performance}
The principal motivation for coarse-graining is the reduction of computational cost while maintaining predictive accuracy. To quantify the efficiency gain achieved by the MLCGP, we compared wall-clock simulation times for atomistic and coarse-grained systems of identical molecular composition. All 64-molecule system simulations were performed on an NVIDIA T4 GPU, while 512-molecule simulations were performed on an NVIDIA A100 GPU. Simulations used LAMMPS for evaluating the dynamics \cite{LAMMPS}. 

Table \ref{tab:timing} summarizes the computational performance of the AA reference and both anisotropic (MLCGP) and isotropic (MLCGP-ISO) coarse-grained models across two system sizes. Both anisotropic and isotropic models achieve substantial and comparable speedups relative to the AA simulation, demonstrating that the anisotropic representation introduces minimal computational overhead. The scaling behavior illustrates that the performance advantage of the coarse-grained representation grows with system size, as larger systems better saturate GPU resources and reduce the relative impact of baseline overhead such as I/O. These results confirm that the proposed anisotropic MLCGP can deliver significant speedups while maintaining quantitative agreement with atomistic structural observables, thereby reaching the central goal of coarse-grained modeling, efficiency without sacrificing accuracy.

\begin{table}[h]
\centering
\caption{Computational performance comparison across models and system sizes. Values represent wall-clock time in ns/day, with speedup factors relative to all-atom (AA) simulations noted.}
\label{tab:timing}
\vspace{0.3cm}
\begin{tabular}{lccc}
\hline
System Size & AA & MLCGP & MLCGP-ISO \\
\hline
64 molecules & 0.196 & 1.343 (6.9$\times$) & 1.389 (7.1$\times$) \\
512 molecules & 0.049 & 1.322 (27$\times$) & 1.431 (29$\times$) \\
\hline
\end{tabular}
\end{table} 

Importantly, water represents a minimal test case with only three atoms per coarse-grained bead. For larger molecules such as polymers, proteins, or complex organic molecules where tens or more atoms are mapped to a single bead, the computational advantage would be substantially greater, as the reduction in DoFs scales directly with the coarse-graining ratio.

\section{Conclusion}
We have presented an anisotropic machine learning coarse-grained potential (MLCGP) capable of accurately capturing both translational and rotational behavior of molecular liquids while achieving considerable computational acceleration relative to fully atomistic simulations. By representing each molecule as an ellipsoidal bead with orientation-dependent features and encoding these features within an E(3)-equivariant message-passing neural network, the model directly learns per-bead energies, forces, and torques from all-atom reference data. The anisotropic representation is physically motivated for systems with asymmetric molecular shapes or directional interactions, where isotropic beads would discard important structural information.

Validation against coarse-grained atomistic trajectories demonstrates that the model reproduces radial and angular distribution functions as well as relative orientation correlations with quantitative accuracy, indicating that both positional and angular interactions are effectively learned. Comparison with an isotropic baseline reveals the benefits of the anisotropic representation, resulting in significant structural correlation improvements over the isotropic model. These results demonstrate that orientation-dependent features are essential for accurate coarse-graining of systems with directional interactions, enabling both correct pair structure and access to rotational observables that are fundamentally inaccessible to spherical bead models, without significant additional computational cost. Additionally, the explicit representation of orientation facilitates the reverse mapping required for active learning workflows, as rotational degrees of freedom constrain atomistic reconstruction in a way that is not available to isotropic models. Computational benchmarks show 7-27$\times$ speedups for small to moderate system sizes, with performance gains increasing for larger simulations. These results confirm that incorporating shape and orientation in MLCGPs provides a practical and necessary route to bridging the gap between atomistic accuracy and mesoscopic simulation efficiency for anisotropic molecular systems.

At the same time, the inclusion of anisotropic features increases the dimensionality of the input space, expanding the data requirements for accurate training. This effect was observed empirically through the volume of additional configurations required during active learning, which exceeded those typical of atomistic potential development. While equivariant architectures help mitigate these challenges, and the performance improvements demonstrated here justify the added cost for systems where molecular anisotropy plays a significant role, sampling efficiency and training set coverage remain critical considerations, particularly for more complex or heterogeneous molecular systems. Similarly, although uncertainty quantification was incorporated to guide the active learning cycle, its performance was unsatisfactory, reflecting ongoing challenges shared with atomistic MLIPs. Developing reliable UQ methodologies and enhanced sampling techniques for active learning will be essential for establishing robust, autonomous pipelines for coarse-grained potential construction.

Looking forward, the framework presented here opens several avenues for future research. While the current work demonstrates clear benefits of anisotropic representations for polar molecules like water, extending this to other molecular classes will help establish when ellipsoidal representations are most valuable versus when simpler isotropic beads suffice. Automated or data-driven mapping strategies are needed to extend the approach to multi-site or heterogeneous molecules without depending on prescribed, heuristic mappings. Integration with hierarchical or multi-scale models, especially within an automated active learning framework, could further enhance the reach of MLCGPs to mesoscale phenomena. Developing MLCGPs directly from first-principles data mediated by all-atom MLIPs or even smaller scale MLCGPs would also provide a systematic path for modeling novel and chemically diverse molecular systems. Finally, coupling active learning with robust UQ has the potential to reduce data requirements and training cost while maintaining accuracy, enabling anisotropic MLCGPs to become general-purpose tools for simulating complex molecular systems at scales currently inaccessible to atomistic methods.

\pagebreak
\printbibliography

\end{document}